# Gravitational Wave Detection in the Introductory Lab

Lior M. Burko[*]
School of Science and Technology, Georgia Gwinnett College,
Lawrenceville, Georgia 30043

February 14, 2016; Revised March 22, 2016

## I. Introduction

A long time ago in a galaxy far, far away two black holes, one of mass 36 solar masses and the other of mass 29 solar masses were dancing their death waltz, leading to their coalescence and the emission of gravitational waves carrying away with them three solar masses of energy. More precisely, it happened 1.3 billion years ago at a distance of 410 Mpc. When the waves were emitted the most complex life forms on Earth were eukaryots. As the gravitational waves propagated toward the Earth it changed much. Five hundred million years after the waves were emitted, or 800 million years ago, the first multicellular life forms emerged on the Earth. The Earth saw the Cambrian explosion 500 million years ago. Sixty-six million years ago the Cretaceous-Paleogene extinction event caused the disappearance of the dinosaurs. The first modern humans appeared 250,000 years ago.

Independently of the event, one hundred years ago, in 1916, Albert Einstein concluded that gravitational waves exist. The construction of LIGO, the Laser Interferometer Gravitational-wave Observatory, commenced 22 years ago. Five years ago Advanced LIGO installation started, and was finished with a full engineering run during September 2015, just in time to catch on September 14, 2015 the gravitational waves that were emitted by the two black holes 1.3 billion years ago. The waves penetrated the southern hemisphere from the general direction of the Magellanic clouds (although one cannot pinpoint the direction with much accuracy because only two detectors were online), traveled through the Earth and were captured from below by the twin LIGO detectors, one in Louisiana near Livingston, and the other in the State of Washington near Hanford, thereby ushering in the new era of gravitational wave astronomy.

---

[*] burko@ggc.edu

## II. What Are Gravitational Waves?

Gravitational waves are ripples in the curvature of space and time, which propagate at the speed of light. Say the source of a gravitational field were to undergo an abrupt change, e.g., say the Sun were to suddenly disappear. How would that affect the orbit of the Earth? Students of the introductory college course learn that the Earth would move uniformly along a tangent to its previous orbit, as required by Newton's First Law. But when will the Earth depart from its orbital motion and start moving uniformly? According to Newtonian gravity that would happen instantaneously, but in reality there must be some delay between the disappearance of the Sun and the moment at which the Earth would start moving uniformly, as the information that the Sun disappeared cannot travel faster than light. In fact, it is thought that the change in the gravitational field propagates as gravitational waves. Gravitational waves are the mechanism by which the changing gravitational field of a changing mass distribution propagates, loosely analogous the electromagnetic waves, which are the mechanism by which a changing electric charge distribution tells remote locations that the source is changed.

The direct detection of gravitational waves did much more than just prove experimentally the concept of gravitational radiation and that binary black hole systems exist, big achievements in themselves; It opened a new window onto the Universe, that many believe will allow us to see hitherto unimagined astrophysical phenomena in addition to many conventional or speculative astrophysical systems. From this point of view, the detection of gravitational waves is arguably even more transformational than the recent 2012 discovery of the Higgs boson and is comparable to the revolution that occurred when Galileo first pointed a telescope to the skies.

When Einstein first predicted the existence of gravitational waves (Einstein, 1916) it was believed that their effect was so miniscule that they would never be discovered, and later their very existence was the topic of a lively debate among relativists (for a history of gravitational waves research see Kennefick, 2007). Both better understanding of astrophysics, specifically about compact objects that move at relativistic speeds and unprecedented progress in technology allowed us, after a century of theory and half a century of experimental searches to finally detect gravitational waves.

**III. How can the detection of gravitational waves be included in the introductory course?**

Recent physics breakthroughs are rarely included in the introductory physics or astronomy course. General relativity and binary black hole coalescence are no different, and can be included in the introductory course only in a very limited sense. (See discussion of gravitational waves in The Physics Teacher in Rubbo *et al*, 2006; Larson et al, 2006; and Spetz, 1984. See also Rubbo *et al*, 2007.) However, we can design activities that directly involve the detection of GW150914, the designation of the gravitation wave signal detected on September 14, 2015 (Abbott *et al*, 2016a), thereby engage the students in this exciting discovery directly. The activities naturally do not include the construction of a detector or the detection of gravitational waves. Instead, we design it to include analysis of the data from GW150914, which includes some interesting analysis activities for students. The same activities can be assigned either as a laboratory exercise in data analysis or as a computational project for the same population of students. The analysis tools used here are simple and available to the intended student population. It does not include the sophisticated analysis tools, which were used by LIGO to carefully analyze the detected signal. However, these simple tools are sufficient to allow the student to get important results.

One may argue that the inclusion of computational projects in the introductory calculus-based course is beneficial to many students, as it allows students to become familiar with a very important tool in the toolkit of the working physicist, in addition to other benefits (Redish and Wilson, 1993; Chabay and Sherwood, 2006). It is the choice of the instructor whether to require a specific computer language (a popular choice is Python), or let the students choose any computer language they already know (or use a spreadsheet such as Excel if they do not know any computer language. In addition, many calculations can be carried out on platforms widely available in the introductory laboratory, such as the PASCO Capstone program.). I am typically in favor of the latter, as I believe that class time in a physics course should not be devoted to teaching programming. However, I have seen many instructors use the former approach with much success. The computational programs I assign in the introductory course (including the present one) can all be done using spreadsheets, such that students without background in programming can still do all the assignments.

When using a spreadsheet the student can calculate a numerical derivative by dividing the difference in the quantity of interest between the time step after the one in question (typically, one cell lower) and the one preceding it (typically, one cell higher), and divide by the same for the time difference. The computation can then be copied for the entire column (skipping the first and last cells, for which the procedure is ill defined), producing the numerical derivative at all time steps (except the first and the last). For example, if Column A includes, say, the data for the time, and column B includes the data for position in the corresponding cells, we can calculate the speed in column C by entering in the cell C2 the formula (B3-B1)/(A3-A1). This formula can then be filled in the remaining of the column.

## IV. Suggestions for the design of class activities

### a) Presentation of the waveform

One can access online the data of the gravitational wave signal and use it for analysis. As these data are somewhat noisy, I prefer to use the numerical relativity waveform, which matches the observed signal very well. One may of course use the actual detected data instead of the numerical relativity waveform, at the cost of having to deal with more noise. The impressive agreement of the numerical relativity waveform and the observed data can be seen in Figure 1. The numerical relativity waveform is a theoretically constructed waveform that was computed using the physical parameters that came out of a detailed computer search of the observed data, looking for a theoretical waveform that best matches the data. Also available online are the residuals, the differences between the numerical relativity waveform and the detected signal. Interested instructors may include the residuals in the activity, or even include a bonus exercise to calculate the cross correlation integral of the detected signal and the numerical relativity waveform.
All data are available at https://losc.ligo.org/events/GW150914/. (All results presented here were obtained from the data file https://losc.ligo.org/s/events/GW150914/GW150914_4_NR_waveform.txt.)

The amount by which a body is distorted (stretched or compressed) relative to its reference length is known as strain. Analogously, the relative stretching or compression of space when a gravitational wave passes through the detector is also called strain. Plotting the strain $h$ as a function of detector time, we obtain Figure 1.

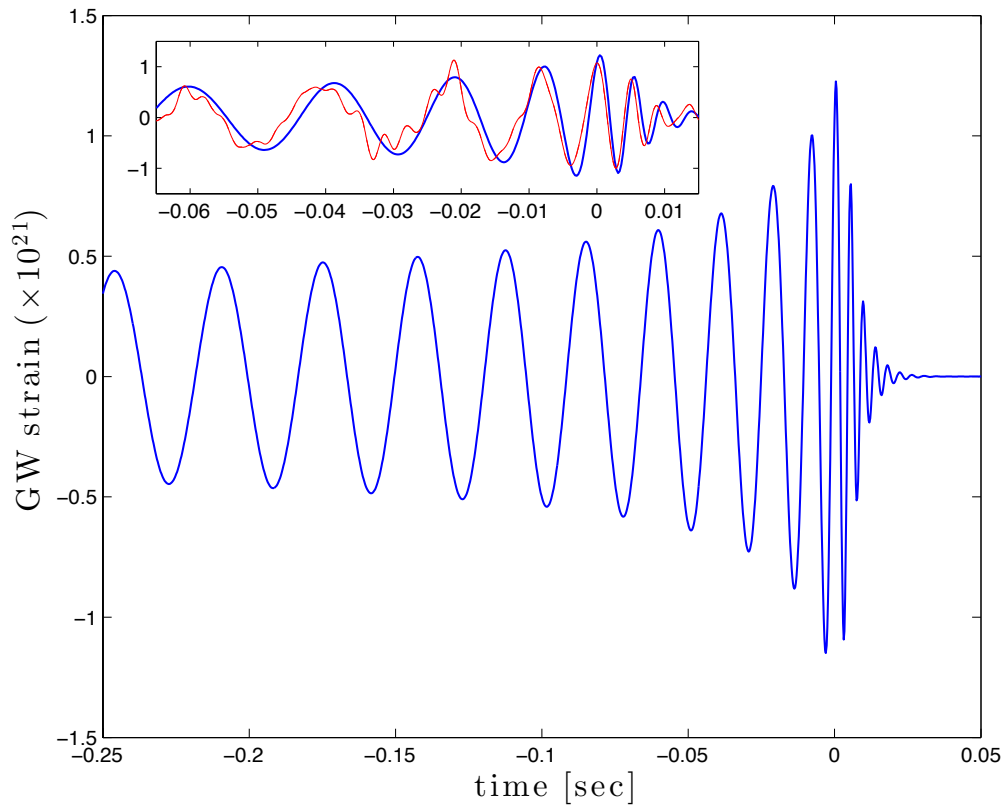

Figure 1: The GW strain $h$ for GW150914 (multiplied by $10^{21}$) as a function of the time $t$ in seconds (from the numerical relativity waveform). The time is shifted so that maximum strain is at $t = 0$ s. The inset shows a section of the data in the main figure (in blue), together with the observed H1 data (in red), i.e., the data from the Hanford detector. The features of the L1 data, the data from the Livingston detector, are similar to those of the H1 data.

The waveform shown in Fig. 1 includes two main parts: first, at earlier times, the so-called "chirp part", and at later parts the ring-down part. The chirp part of the waveform is so named because both the amplitude and frequency increase as a function of time, the features of a bird's chirp. While students of the introductory course see oscillations with varying amplitude (say, in the context of damped harmonic oscillations), they do not normally see varying frequencies. This chirp part of the waveform can be a good laboratory or computational exercise in analyzing data with a varying frequency, specifically the emphasis on a local calculation of frequency as opposed to a global calculation. It is instructive to let the students separate the signal into its two main parts, and emphasize that each part has different properties and therefore requires different analysis tools.

**b) Basic relations satisfied by gravitational waves from a coalescing binary**

For the analysis of the gravitational wave signal we will need some basic relations that the parameters of the waves satisfy. The following discussion is a justification of these relations based on scaling arguments. The full equations require a detailed solution using general relativity, which is beyond the scope of this Paper and the level of the intended audience.

Consider a system of total mass $M$ and of typical size $R$, whose moment of inertia scales like $MR^2$. An unchanging mass distribution cannot radiate. Gravitational waves are therefore generated by a time-changing moment of inertia. A uniformly moving mass also cannot radiate, because for a co-moving observer it is seen as static. As the presence of radiation must be observer independent, no other inertial observer can measure any radiation either. One must therefore have an accelerating mass distribution to have radiation. More specifically, the waves depend on the second time derivative of the moment of inertia. (An analogous argument applies also for accelerating electric charges that emit electromagnetic waves.)

The luminosity of any wave depends on the square of the amplitude. (The energy of the wave must be positive even though the wave oscillates between positive and negative values, and therefore must depend on an even power of the amplitude). Conservation of energy then dictates that the luminosity drops with distance as the square of the distance, and therefore the amplitude must decay like the reciprocal of the distance to the observation point, $r$. In what follows we consider only the scaling relations and do not worry about constant or dimensionless factors.

Replacing a time derivative by the frequency $f$, we find that $h \sim MR^2 f^2/r$. Using Kepler's third law for a binary system, or $f \sim (M/R^3)^{1/2}$, we find that $h \sim \frac{M^2}{rR}$. In fact, this scaling relation is conventionally written (after using Kepler's law again and introducing the chirp mass) as $h \sim \frac{1}{r} \mathcal{M}_C^{5/3} f^{2/3}$. When the constant factors are worked out, the amplitude relation becomes

$$h = \frac{1}{c^2} \frac{1}{D_L} \left(\frac{G}{c^2} \pi f\right)^{2/3} (G\mathcal{M}_C)^{5/3}, \tag{1}$$

where we write the distance between the detector and the source as the luminosity distance $D_L$, and $f$ is the frequency of the waves. Here, $c$ is the speed of light and $G$ is Newton's constant. Simple arguments for the amplitude equation in terms of dimensional analysis are given in Schutz (2003).

Next, we start with the luminosity $L \sim 4\pi r^2 f^2 h^2$, in a similar way to how luminosity of light is written. (The luminosity must depend on an even power of $f$ because the energy of the waves must be positive for both positive and negative frequencies. A detailed calculation shows that the luminosity depends on the square of the frequency.) Using Kepler's law and the amplitude relation, we find that $L \sim M^5/R^5$. On the other hand, the luminosity $L$ equals (the negative of) the rate of change of the total energy of the source. The total energy is the gravitational potential energy, or $E \sim -M^2/R$, and therefore $\dot{E} \sim \frac{M^2}{R^2} \dot{R}$. Requiring that $L = -\dot{E}$ we find that $\dot{R} \sim -\frac{M^3}{R^3}$.

Differentiating Kepler's third law, $f \sim (M/R^3)^{1/2}$, the rate of change of the frequency $\dot{f} \sim \frac{M^{1/2}}{R^{5/2}} \dot{R}$. Substituting our result for $\dot{R}$, we find that $\dot{f} \sim \frac{M^{7/2}}{R^{11/2}}$. Using Kepler's law again, and reintroducing the chirp mass, we write our result as $\dot{f} \sim \mathcal{M}_C^{5/3} f^{11/3}$. When the constant factors are included, we find that

$$\dot{f} = \frac{96}{5} \frac{\pi^{8/3}}{c^5} f^{11/3} (G\mathcal{M}_C)^{5/3}. \tag{2}$$

This equation for the rate of change of the frequency is known as the chirp.

### c) Analysis of the data

Armed with Eqs. (1) and (2), we can now start analyzing the waves.

i) Finding the frequency

Assume that the chirp part of the waves has the form
$$h(t) = A(t) \sin(\omega(t) \times t),$$
where both the amplitude $A$ and the angular frequency $\omega$ are functions of the time $t$. A simple analysis tool is to assume that the changes in the parameters $A$ and $\omega$ are slow compared with the measurement time step. That is, we may assume that as long as the change in the parameter is slow on the scale of the time increment, at each value of the time we can treat the waveform as if it had constant parameters in the first approximation.

Therefore, we can differentiate the strain expression above twice in this approximation to get
$$\omega^2 = -\frac{\ddot{h}}{h}$$
where an overdot denotes the time derivative. Students use this relation to get a numerical estimate of angular frequency and how it changes with time. Notice that we do not need to find a phase constant, so it is assumed without loss of generality to equal zero. To obtain the second derivative students can either write a short code that calculates the numerical derivative for a computational project, use a spreadsheet such as EXCEL as described above, or use software commonly available in the introductory laboratory such as PASCO Capstone. The latter allows finding the "acceleration" of an oscillating signal.

In Fig. 2 we plot the angular frequency as a function of the time, after smoothing out the noise with a simple averaging algorithm. In order to average out noise the student can add an averaging routine to the code, or add a column to the EXCEL spreadsheet that finds the simple average for a cell for a certain range of cells around it. If using Capstone, the student can use a best fit curve as the noise smoothing algorithm or use the built-in curve smoothing function. The averaging done here is for intervals of 25ms. The unsmoothed data for the angular frequency may include isolated points at which the values are very different than in neighboring points. Such points may be removed from the data set, which makes the smoothing process simpler. In what follows, all our calculations are done in the detector frame. (The two black hole masses given in the first paragraph are given in the source frame.) We therefore compare with the LIGO values for the detector frame.

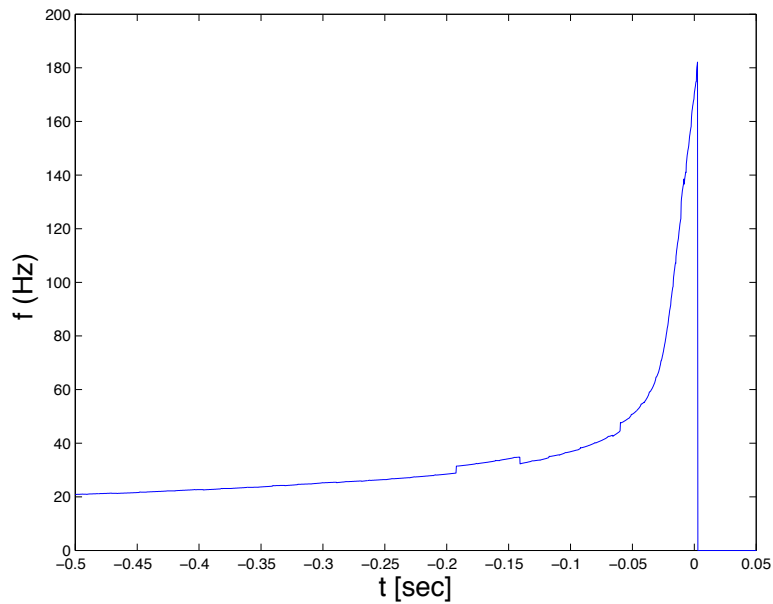

Figure 2: Frequency as a function of the time. The increasing frequency is shown from about 20 Hz to over 150 Hz during 8 cycles of the binary.

ii) The Chirp mass

The leading-order dependence of the waveform evolution (i.e., the evolution in the early parts of the waveform) on the masses of the binary black holes is through the so-called chirp mass. Specifically, the independent masses of the binary black holes cannot be easily deduced, as more information (which includes relativistic effects which are beyond the scope of the present Paper) is needed in order to find the individual masses. It is called the chirp mass because it is the parameter responsible for the chirp-like sound (increasing frequency and amplitude) that is heard when a gravitational wave signal from a binary merger is converted to sound waves. The chirp mass is defined as

$$\mathcal{M}_C = \frac{(m_1 m_2)^{3/5}}{(m_1 + m_2)^{1/5}},$$

where $m_1$ and $m_2$ are the masses of the two members of the binary. It is this particular combination of the individual masses that can be found from the leading order evolution of $h(t)$. As the waveform depends at first only on the chirp mass, the waveform does not depend on any other combination of the masses, and for that reason it is a useful parameter to introduce. The frequency evolution of the wave, $\dot{f}$, can be used from Eq. (2). This equation can next be inverted to give the chirp mass in terms of the calculated $f$ and $\dot{f}$, which can be constructed from the GW150914 data. (That is, having found $f(t)$ the student can now calculate the numerical derivative to find $\dot{f}$.)

In Fig. 3 we plot the chirp mass, after some smoothing done as above. Notice that the chirp mass is well defined using these simple expressions only during the early part of the inspiral. We therefore read the chirp mass as about 30 solar masses. In finding the chirp mass one can again either write a short numerical code, use a spreadsheet, or alternatively use PASCO Capstone or a similar program. At early times the chirp mass is nearly constant, but then starts oscillating wildly. Indeed, our calculation of the chirp mass is only intended to be accurate in the early parts of the inspiral, when the two black holes are still far from each other. As they get closer relativistic corrections become important, and the simple relations that we use become less and less accurate approximations, as the behavior of the chirp mass shows.

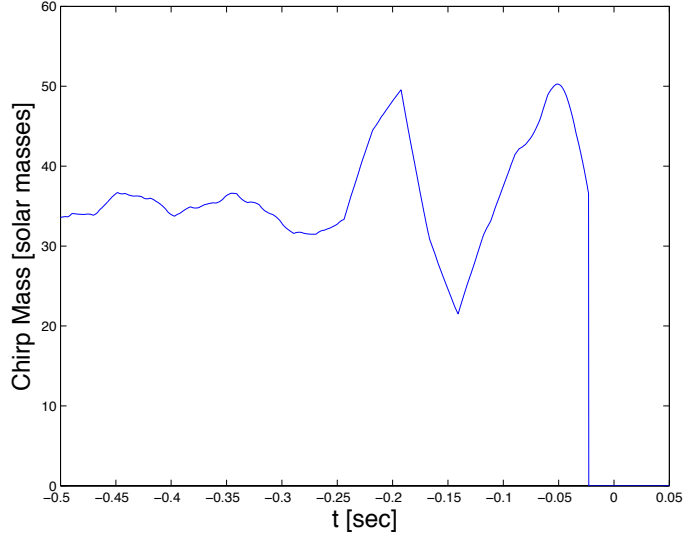

Figure 3: The chirp mass as a function of time. We only read the chirp mass from the early part of the evolution. In the early part of the merger the chirp mass is nearly constant, and it starts oscillating widely after relativistic corrections become important.

iii) Evaluating the total mass

The total mass of the binary black hole system can be crudely estimated from the chirp mass. If we assume that the two black holes have mass ratio $q = m_1/m_2$, one can use the definition of the chirp mass and solve for the total mass: First substitute $m_1 = m_2 q$ in the definition for the chirp mass, and find that $m_2 = \left[(1+q)^{1/5}/q^{3/5}\right]\mathcal{M}_C$. Next, write $M = m_1 + m_2 = (1+q)m_2$, and find that the total mass $M = \frac{(1+q)^{6/5}}{q^{3/5}}\mathcal{M}_C > 2^{6/5}\mathcal{M}_C$. Notice that one does not need to know the mass ratio in order to find this inequality (as $\frac{(1+q)^{6/5}}{q^{3/5}}$ as a function of $q$ is bounded from below by $2^{6/5}$), even though the mass ratio is needed in order to find the total mass itself (and not just a lower bound). However, determining the mass ratio requires relativistic effects that are beyond the scope of this Paper. Note that the lower bound is achieved when the two masses are equal, and for unequal masses the total mass is always greater. Specifically, $M > \sim 70$ solar masses.

iv) Determination of the mass and spin rate of the final black hole

Next, we turn attention to the ring-down part of the waves. Inspection reveals it is a constant frequency signal of exponentially decaying amplitude. A good question to ask the students is what kind of system does a signal such as that suggest. It may be useful to refrain until this point from calling it the "ring-down" part, as the name suggests the answer to the question. If such a question is posed, instructors can divide the signal into the early part and late part, say, instead of the chirp part and ring-down part when the signal is first introduced.

When a bell rings it emits sound in a characteristic spectrum of frequencies. An ideal bell, which rings on forever with no losses, would emit sound in frequencies known as its normal modes. Any real bell rings down because of the loss of energy to sound waves, and its frequencies are known as quasi-normal modes. Black holes are like bells: when perturbed, they too emit radiation in the form of gravitational waves in frequencies of their quasi-normal modes. In fact, black holes are poor bells, as they don't get to ring many times before they settle down to a stationary, quiescent state. The spectrum of black hole quasi-normal modes is complex, and the frequencies depend only on the mass and spin angular momentum of the black hole. In order to prove that the source is a black hole, multiple quasi-normal mode frequencies need to be measured. However, if we assume that the ring-down phase of the radiation is given by the slowest decaying quasi-normal mode of a spinning black hole that is perturbed from its stationary state, we can calculate the mass and spin angular momentum of the final black hole. As we can observe in the data only one mode (that is, one frequency and one time constant for the decay rate), we assume that this mode is the slowest damped mode of the black hole's quasi-normal spectrum.

The chirp waveform is followed by the quasi-normal ring-down of the final single black hole. When the final black hole is first created it is highly distorted, and emits gravitational waves that reduce this distortion until it get to its stationary and quiescent state. Indeed, the waveform for GW150914 includes an exponential decay of the signal with constant frequency after the merger. (In practice, the parameters for the ring-down published in Abbott, B.P. *et al* (2016a) are found from the numerical relativity waveform, as we do here.) In Fig. 4 we show the part of the signal after the merger (positive values of time). This is a good example of the benefit in plotting a semi-logarithmic graph, as the straight envelope shows the exponential decay of the signal, and the constant frequency is easily visible. The frequency and decay rate can be read from the graph, or instead can be determined from the fit function of programs such as PASCO Capstone. Specifically, fitting to a damped sine function of the form $h(t) = Ae^{-t/\tau}\sin(2\pi f t + \varphi_0)$ we determine the parameters to be $\tau = 4.4 \times 10^{-3}$ sec, and $f = 236$ Hz. These values agree nicely with the values of the LIGO experiment, $\tau = (4.0 \pm 0.3) \times 10^{-3}$ sec, and $f = 251 \pm 8$ Hz (Abbott *et al*, 2016b). Figure 5 shows the best fit of the data for the ring-down portion of the wave, which we did using PASCO Capstone. A more sophisticated approach, also available in PASCO Capstone, involves the Fast Fourier Transform of the decaying oscillations, but is generally too sophisticated for the intended student population. As the ring-down is assumed here to be a monochromatic signal (single mode), the simpler fitting to a decaying sine function is sufficient.

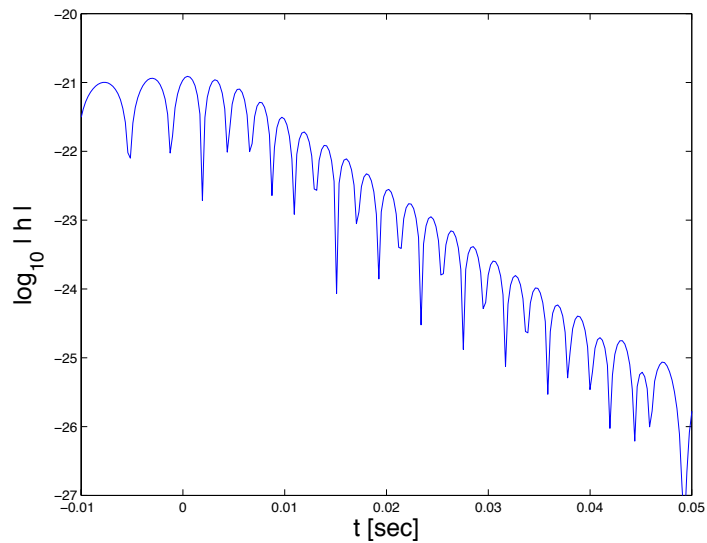

Figure 4: A plot of the logarithm of the signal as a function of time for the ring-down part of the waveform. The exponential decay rate and the constant frequency are visible in the semi-logarithmic representation.

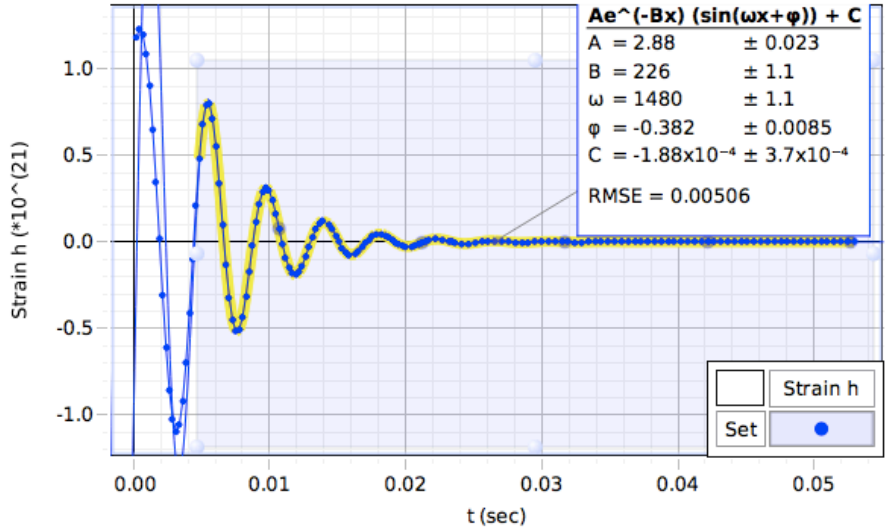

Figure 5: Best fit for the parameters of a damped sine function for the ring-down part of the GW150914 signal.

The frequency $f$ and time constant $\tau$ can next be used to find the mass $M_f$ and specific angular momentum $a$ of the final black hole. Here, $a = cJ/(GM^2)$, where $J$ is the black hole's spin angular momentum. The specific angular momentum $0 \leq a \leq 1$, and is a convenient dimensionless variable. We use here a simple method that is based on simple fit functions for the black hole's quasi-normal modes. Specifically, let the quality factor $Q = \pi f \tau$. Then,

$$\frac{G}{c^3} \times 2\pi f M_f = f_1 + f_2(1-a)^{f_3}$$

and

$$Q = q_1 + q_2(1-a)^{q_3},$$

where the fit parameters for the principal least damped mode are $f_1 = 1.525, f_2 = -1.157, f_3 = 0.129$, and $q_1 = 0.700, q_2 = 1.419, q_3 = -0.499$ (Berti *et al*, 2009). These fit equations were developed to reproduce the frequency and quality factor to better than 1% with the least number of parameters for the expected range of the specific angular momentum. These fit equations are not derived from physical principles, and are not even a phenomenological fit. Instead, they are the simplest mathematical equations that reproduce the known data to a given accuracy level. (The values that the equations fit, however, are derived from a clear physical model for the quasi-normal ring-down.) While this may perhaps not provide some students with a strong physical motivation for using these fit equations (and indeed, one should prefer to be able to physically motivate equations that one uses), it is perhaps an opportunity to point out that when one does research, sometimes one does what one can do. Instructors, who feel that the lack of physical motivation for the equations and the fit parameters are reasons to not use them, may omit this part of the discussion, and have the students only find the frequency and time constant of the ring-down. These instructors can then discuss how the mass and spin of the final black hole can be found from data available in tables, which were derived from solid physical models. The author's experience with his own students is that finding the mass and spin of the final black hole is one of the favorite parts for many students.

Be that as it may, these equations are, however, extremely effective for our purposes. We can now solve this set of two equations for the two unknown $M_f$ and $a$, and find that $M_f = 67\ M_{Sun}$ and $a = 0.67$. In this calculation we use $G/c^3 = 0.248\times10^{-35}$ s kg$^{-1}$, and $M_{Sun} = 1.99\times10^{30}$ kg. The values reported in the LIGO paper are $M_f = 67 \pm 4\ M_{Sun}$ and $a = 0.67^{+0.05}_{-0.07}$. Since we found above that the total mass before the merger was $>\sim 70\ M_{Sun}$, we infer that at least 3 $M_{Sun}$ were radiated in the form of gravitational waves. Taking the merger time to be 0.2s, the average power emitted is $3\times10^{55}$ erg s$^{-1}$.

v) Distance to the source

If this activity is assigned to students of an introductory astronomy or cosmology course, where the concepts of luminosity distance and cosmic expansion are introduced, the activity can be extended to find the luminosity distance. For students of the introductory physics course the calculation of the luminosity distanced can be skipped. One starts with the amplitude equation (1), and combines it with the frequency evolution equation (2). One can then solve for $D_L$ in terms of only quantities that are directly measurable from the waveform, specifically,

$$D_L = \frac{5}{96\pi^2}\frac{c}{h}\frac{\dot{f}}{f^3}.$$

The last result is an extremely important one, as it shows that the distance from the source can be measured directly from the waveform. This is a unique case for astronomy, as the distance from a source observed with light cannot be inferred from only the signal, and other methods need to be used in order to estimate the distance (the so-called distance ladder). E.g., there was at first a fierce debate about the distance to quasars before their redshift was measured, as the distance could not be inferred directly from the observations. The distance to the source of GW150914, 400 MPc, is much longer than the typical distance between galaxies, ~1 Mpc, which itself is about 50 times the size of a single galaxy. If the students are familiar with cosmological expansion and know the redshift–distance relation, they can check that the source is at a cosmological distance, specifically the redshift $z = 0.1$.

In summary, the detection of gravitational waves with LIGO provides physics or astronomy instructors with an opportunity to engage students of the introductory course with this very exciting and important discovery using simple data analysis tools that are available to the students.

This research has made use of data, software and/or web tools obtained from the LIGO Open Science Center (https://losc.ligo.org), a service of LIGO Laboratory and the LIGO Scientific Collaboration. LIGO is funded by the U.S. National Science Foundation.

# Erratum: Gravitational Wave Detection in the Introductory Lab
## [The Physics Teacher **55**, 288 (2017)]

Lior M. Burko, Georgia Gwinnett College

December 6, 2018

There is a typo in the amplitude relation, Eq. (1). The correct equation is

$$h = \frac{2}{c^4} \frac{1}{D_L} (\pi f)^{2/3} (GM_c)^{5/3}.$$

Notice that this correction involves only constant factors, and therefore the derivation, which was done using scaling relations, remains unaffected.